\documentclass[12pt]{article}
\usepackage[english,german,french,polish]{babel}
\usepackage[T1]{fontenc}

\begin{document}

\baselineskip 0.72cm
\topmargin -0.4in
\oddsidemargin -0.1in

\let\ni=\noindent

\renewcommand{\thefootnote}{\fnsymbol{footnote}}

\newcommand{\CKM}{Cabibbo-Kobayashi-Maskawa }

\newcommand{\Stk}{SuperKamiokande }

\newcommand{\SM}{Standard Model }

\pagestyle {plain}

\setcounter{page}{1}

\pagestyle{empty}




~~~~~
\begin{flushright}
IFT-02/16
\end{flushright}

\vspace{0.2cm}

{\large\centerline{\bf Option of light righthanded neutrinos avoiding oscillations{\footnote {Work supported in part by the Polish State Committee for Scientific Research (KBN), grant 5 P03B 119 20 (2001--2002).}}}}

\vspace{0.5cm}

{\centerline {\sc Wojciech Kr\'{o}likowski}}

\vspace{0.23cm}

{\centerline {\it Institute of Theoretical Physics, Warsaw University}}

{\centerline {\it Ho\.{z}a 69,~~PL--00--681 Warszawa, ~Poland}}

\vspace{0.5cm}

{\centerline{\bf Abstract}}

\vspace{0.3cm}

A new explicit model of neutrino texture is presented, where in the overall $6\times 6$ mass matrix  the lefthanded and righthanded Majorana $3\times 3$ components are diagonal with equal entries of opposite sign, while the Dirac $3\times 3$ component is given as a diagonal hierarchical structure (possibly similar to the charged lepton and quark $3\times 3$ mass matrices) deformed by the popular nearly bimaximal $3\times 3$ mixing matrix. Then, all neutrino masses are light and $ m_1 = -m_4 $, $m_2 = -m_5 $, $m_3 = -m_6 $. The resulting neutrino oscillation formulae are {\it identical} with those working in the effective texture of  three active  neutrinos, based on this nearly bimaximal mixing matrix. Three (conventional) sterile neutrinos {\it do not oscillate} and so, are strictly {\it decoupled}. The suggested LSND effect {\it vanishes}. The not observed Chooz oscillation effect is consistently negligible.

\vspace{0.4cm}

\ni PACS numbers: 12.15.Ff , 14.60.Pq , 12.15.Hh .

\vspace{0.8cm}

\ni May 2002

\vfill\eject

~~~~~
\pagestyle {plain}

\setcounter{page}{1}
 
\ni {\bf 1. Introduction.} A basic theoretical question, induced by the recent developments in neutrino physics implying massive neutrinos, concerns the fate of neutrino righthanded components $\nu_{\alpha R} \;(\alpha = e,\, \mu,\,\tau)$, while the lefthanded components $\nu_{\alpha L} \;(\alpha = e,\, \mu,\,\tau)$ appear in the \SM as active neutrinos. This is equivalent to the question of the role in Nature of (conventional) sterile neutrinos $\nu_{\alpha R} $ and $(\nu_{\alpha R})^c$, where $(\nu_{\alpha R})^c \neq \nu_{\alpha L}$ due to the opposite lepton number $L$ of both sides. The lepton  number, in fact, is experimentally well defined, though it is not expected to be strictly conserved. In the present note, the (conventional) sterile neutrinos $\nu_{\alpha R} $ and $(\nu_{\alpha R})^c$ will be introduced through the generic neutrino mass term (11). First, however, we will consider the effective neutrino-mass term (3) of Majorana type, where only the active neutrinos $\nu_{\alpha L}$ and  $(\nu_{\alpha L})^c$ appear. 

As is well known, the popular nearly bimaximal mixing matrix for three active neutrinos $\nu_{e L}$, $\nu_{\mu L}$, $\nu_{\tau L}$ [1],

\begin{equation} 
U^{(3)} =   \left( \begin{array}{ccc} c_{12} & s_{12} & 0 \\ -s_{12}c_{23} & c_{12}c_{23} & s_{23} \\ s_{12}s_{23} & -c_{12}s_{23} & c_{23} \end{array} \right) \; ,
\end{equation} 

\ni arises from its generic form {\it \`{a} la} \CKM [2] by putting 

\begin{equation}
s_{13} = 0 \;\;,\;\; c_{12} \simeq \frac{1}{\sqrt{2}} \simeq s_{12}\;\;,\;\; c_{23} \simeq \frac{1}{\sqrt{2}} \simeq s_{23} \;.
\end{equation}

\ni Here, $c_{ij} =\cos \theta_{ij}$ and $s_{ij} =\sin \theta_{ij}$. Such a neutrino mixing matrix is globally consistent with oscillation experiments [3] for solar $\nu_e$'s and atmospheric $\nu_\mu$'s as well as with the negative Chooz experiment for reactor $\bar{\nu}_e$'s. It cannot explain, however, the possible LSND effect for accelerator $\bar{\nu}_\mu$'s (and $\nu_\mu$'s) that, if confirmed, may require the existence of one at least, extra (sterile) light neutrino $\nu_{s L}$ (different in general from $(\nu_{\alpha R})^c$). 

If the active neutrinos $\nu_{\alpha L}$ are of Majorana type, their effective mass term in the Lagrangian has the form

\begin{equation} 
- {\cal L}^{(3)}_{\rm mass} = \frac{1}{2}\sum_{\alpha \beta} \overline{(\nu_{\alpha L})^c} M^{(3)}_{\alpha \beta} \nu_{\beta L} + {\rm h.\,c.} \;,
\end{equation} 

\ni where the mass matrix $M^{(3)} = \left( M^{(3)}_{\alpha \beta} \right)$ is symmetric due to the identity $\overline{\nu_{\alpha L}} (\nu_{\beta L})^c = \overline{\nu_{\beta L}} (\nu_{\alpha L})^c $ (here, the normal ordering of bilinear neutrino terms is implicit). In this case, $(\nu_{\alpha L})^c$  behaves as $\nu_{\alpha R} $, though their lepton number $L$ is opposite. This number is not conserved, of course, in the mass term (3) inducing the changes $\Delta L = \pm 2$.

In the flavor representation, where the charged-lepton $3\times 3$ mass matrix is diagonal, the mixing matrix $U^{(3)}$, when multiplied from the right  by the Majorana phase matrix diag$(1 ,\,e^{i\rho} ,\,e^{i\sigma})$ in the case of Majorana-type neutrinos, becomes the diagonalizing matrix that transforms the complex symmetric neutrino $3\times 3$ mass matrix $M^{(3)}$ into the matrix diag$(m_1\,,\,m_2\,,\,m_3)$. Here $ m_1 \leq m_2 \leq m_3 $ are nonnegative neutrino masses. This is equi\-valent to the following complex orthogonal transformation [4]:

\begin{equation} 
U^{(3)\,T} M^{(3)} U^{(3)} = {\rm diag}(m_1 \,,\,m_2\, {\rm e}^{-2 i \rho} \,,\,m_3\, {\rm e}^{-2 i \sigma}) \;\,,\,\; U^{(3)\,T} U^{(3)*} = 1 = U^{(3)*} U^{(3)\,T}\;.
\end{equation} 

\ni The reverse transformation reads 

\begin{equation} 
M^{(3)} = U^{(3)*} {\rm diag}(m_1 \,,\,m_2 {\rm e}^{-2 i \rho} \,,\,m_3 {\rm e}^{-2 i \sigma}) U^{(3) \dagger} \;.
\end{equation} 

\ni However, even in the Majorana case, the flavor and mass active neutrinos, $\nu_{\alpha L}\;(\alpha = e, \mu, \tau)$ and $\nu_{i L}\;(i = 1,2,3)$, are related through the unitary transformation

\begin{equation} 
\nu_{\alpha L}  = \sum_i U^{(3)}_{\alpha i}  \nu_{i L} \;\;,\;\; U^{(3)\,\dagger} U^{(3)} = 1 = U^{(3)} U^{(3)\,\dagger}\;,
\end{equation} 

\ni where $ U^{(3)} = \left(U^{(3)}_{\alpha i}\right)$. The unknown Majorana phases $\rho $ and $\sigma $ as well as the Dirac phase $\delta $, the latter appearing in the generic form of $ U^{(3)}$ if $ s_{13} \neq 0$ ({\it e.g.} through $ U_{e 3} = s_{13} e^{-i\delta}$), may lead to CP violation. Note, however, that CP violation in the neutrino oscillations may be caused only by $\delta $.

The rate of the neutrinoless double $\beta $ decay (allowed only in the case of Majorana-type $\nu_{e L}$) is proportional to $ m^2_{e e}$, where $ m_{e e} \equiv |M^{(3)}_{e e}|$ with $M^{(3)}_{e e}$ as given in Eq. (5) for $ \alpha = e$ and $ \beta = e$. The suggested experimental upper limit of $m_{e e}$ is $m_{e e} \stackrel{<}{\sim} (0.35 - 1)$ eV [4]. If $\rho = 0 $ and $\sigma = 0 $, then $|M^{(3)}_{e e}| = |\sum_i U^{(3)\,2}_{e i} m_i | = c^2_{12} m_1 + s^2_{12} m_2\,$, where the form (1) for $U^{(3)}$ is used.

The familiar neutrino-oscillation formulae, valid in the case of $ U_{\alpha i}^{(3)*} = U^{(3)}_{\alpha i}$ (where the possible CP violation is ignored), are

\begin{equation} 
P(\nu_\alpha \rightarrow \nu_\beta) = |\langle \nu_{\beta L}| e^{i PL} |\nu_{\alpha L} \rangle |^2 = \delta _{\beta \alpha} - 4\sum_{j>i} U^{(3)}_{\beta j} U^{(3)}_{\beta i} U^{(3)}_{\alpha j} U^{(3)}_{\alpha i} \sin^2 x_{ji} 
\end{equation}

\ni with

\begin{equation} 
x_{ji} = 1.27 \frac{\Delta m^2_{ji} L}{E} \;,\; \Delta m^2_{ji}  = m^2_j - m^2_i \;,\, p_i \simeq E -\frac{m^2_i}{2E} 
\end{equation} 

\ni ($\Delta m^2$, $L$ and $E$ are measured in eV$^2$, km and GeV, respectively). Here, $P(\nu_\alpha \rightarrow \nu_\beta) = P(\bar{\nu}_\alpha \rightarrow \bar{\nu}_\beta) = P(\nu_\beta \rightarrow \nu_\alpha)$, when CP violation in neutrino oscillation is ignored (and CPT theorem used). For the mixing matrix $ U^{(3)}$ as given in Eq. (1), the formulae (7) lead, in particular, to the following oscillation probabilities:

\begin{eqnarray} 
P(\nu_e \rightarrow \nu_e)\, & = & 1 - (2c_{12}s_{12})^2 \sin^2 x _{21} \;, \nonumber \\
P(\nu_\mu \rightarrow \nu_\mu) & = & 1 - (2c_{12}s_{12})^2 c_{23}^4 \sin^2 x _{21} - (2c_{23}s_{23})^2 (s_{12}^2\sin^2 x_{31} + c_{12}^2 \sin^2 x_{32}) \nonumber \\ & \simeq & 1 - (2c_{12}s_{12})^2 c_{23}^4 \sin^2 x _{21} - (2c_{23}s_{23})^2 \sin^2 x_{32}  \;, \nonumber \\ 
P(\nu_\mu \rightarrow \nu_e) & = & (2c_{12}s_{12})^2 c_{23}^2 \sin^2 x _{21} \;, 
\end{eqnarray} 

\ni where the second step for $P(\nu_\mu \rightarrow \nu_\mu)$ works in the case of $\Delta m^2_{31} \simeq \Delta m^2_{32}$. The remaining oscillation probabilities, $P(\nu_\mu \rightarrow \nu_e)$, $P(\nu_\mu \rightarrow \nu_\tau)$ and $P(\nu_\tau \rightarrow \nu_\tau)$, follow already from Eqs. (9) through the probability sum rules $\sum_\beta P(\nu_\alpha \rightarrow \nu_\beta) = 1 $. The relations (9) imply that

\begin{eqnarray} 
P(\nu_e \rightarrow \nu_e)_{\rm sol}\;\;\;\, & = & 1 - (2c_{12}s_{12})^2 \sin^2 (x _{21})_{\rm sol}\;, \nonumber \\
P(\nu_\mu \rightarrow \nu_\mu)_{\rm atm} \;\; & \simeq & 1 - (2c_{23}s_{23})^2 \sin^2 (x _{32})_{\rm atm} \;, \nonumber \\
P(\bar{\nu}_\mu \rightarrow \bar{\nu}_e)_{\rm LSND} & = & (2c_{12}s_{12})^2 c_{23}^2 \sin^2 (x _{21})_{\rm LSND} \simeq 0 \;, \nonumber \\ 
P(\bar{\nu}_e \rightarrow \bar{\nu}_e)_{\rm Chooz} & = & 1 - (2c_{12}s_{12})^2 \sin^2 (x _{21})_{\rm Chooz} \simeq 1\;, 
\end{eqnarray} 

\ni since $(x _{21})_{\rm atm} \ll (x _{32})_{\rm atm} = O(1)$, $(x _{21})_{\rm LSND} \ll (x_{21})_{\rm sol} = O(1)$ and $(x _{21})_{\rm Chooz} \simeq (x_{21})_{\rm atm} \ll O(1)$, the first inequality being valid in the case of $\Delta m^2_{21} \ll \Delta m^2_{32} \simeq \Delta m^2_{31}$.  The labels sol, atm, LSND and Chooz refer to the conditions of solar, atmospheric, LSND and Chooz neutrino experiments.

Experimental estimations for solar $\nu_e$'s and atmospheric $\nu_\mu$'s  are $\theta_{12} \sim 32^\circ$, $\Delta m^2_{21} \sim 5\times 10^{-5}\;{\rm eV}^2$ [5] and $\theta_{32} \sim 45^\circ$, $\Delta m^2_{32} \sim 2.5\times 10^{-3}\;{\rm eV}^2$ [6], respectively (they are best-fit values; for solar $\nu_e$'s they correspond to the MSW Large Mixing Angle solution that seems to be optimal). The above angles imply $c_{12} \sim 1.2/\sqrt{2}$, $s_{12} \sim 0.75/\sqrt{2}$ and $c_{23} \sim 1/\sqrt{2} \sim s_{23}$, thus deviations from maximal mixing for solar $\nu_e$'s are considerable.

According to the popular viewpoint, the active-neutrino effective mass term (3) arises through the familiar see-saw mechanism from the generic neutrino mass term

\begin{equation} 
- {\cal L}_{\rm mass} = \frac{1}{2} \sum_{\alpha \beta} \left( \overline{(\nu_{\alpha L})^c} \,,\, \overline{\nu_{\alpha R}}\right) \left( \begin{array}{cc} M^{(L)}_{\alpha \beta} & M^{(D)}_{\alpha \beta} \\ M^{(D)}_{\beta \alpha} & M^{(R)}_{\alpha \beta} \end{array} \right) \left( \begin{array}{c} \nu_{\beta L} \\ (\nu_{\beta R})^c \end{array} \right) + {\rm h.\,c.} 
\end{equation} 

\ni including both the active neutrinos $\nu_{\alpha L}$ and $(\nu_{\alpha L})^c$ as well as the (conventional) sterile neutrinos $\nu_{\alpha R} \neq (\nu_{\alpha L})^c$ and $(\nu_{\alpha R})^c \neq \nu_{\alpha L} \;(\alpha = e\,,\,\mu\,,\,\tau)$. In the see-saw case, the Majo\-rana righthanded mass matrix $ M^{(R)} = \left(M^{(R)}_{\alpha \beta}\right)$ is presumed to dominate over the Dirac  mass matrix $ M^{(D)} = \left(M^{(D)}_{\alpha \beta}\right)$ that in turn is expected to dominate over the Majorana lefthanded mass matrix $ M^{(L)} = \left(M^{(L)}_{\alpha \beta}\right)$ (the latter may be even zero). Such a mechanism leads effectively to the active-neutrino mass matrix $ M^{(3)} = \left(M^{(3)}_{\alpha \beta}\right)$ appearing in the mass term (3). Then, $ M^{(3)} \simeq - M^{(D)}  M^{(R)\,-1}M^{(D)\,T}$, while the (conventional) sterile neutrinos get approximately $M^{(R)}$ as their effective mass matrix and so, are practically {\it decoupled} from the active neutrinos. Thus, the assumption of dominance of $M^{(R)}$ over $M^{(D)}$ (and $M^{(D)}$ over $M^{(L)}$) guarantees here the desired smallness of $m_1\,,\,m_2\,,\,m_3$. This approach is consistent with the GUT viewpoint on the massive-neutrino unification.

In the present note, we study a new explicit model for the overall $6\times 6$ mass matrix
 
\begin{equation}
M = \left( \begin{array}{cc} M^{(L)} & M^{(D)} \\ M^{(D)\,T}  & M^{(R)} \end{array} \right) 
\end{equation}

\ni appearing in the generic neutrino mass term (11). Now, $M^{(L)}$ and $M^{(R)}$ get the same magnitude (but opposite sign). The detailed model gives {\it exactly the same} neutrino oscillation formulae (9) and (10) as the previous effective model based on the nearly bimaximal $3\times 3$ mixing matrix (1). The (conventional) sterile neutrinos $(\nu_{\alpha R})^c$ {\it do not oscillate} with the active neutrinos $\nu_{\alpha L}$ (and with themselves) and so, are strictly {\it decoupled} from the active neutrinos (because of different reasons, than those causing practical decoupling of $(\nu_{\alpha R})^c$ from $\nu_{\alpha L}$ in the see-saw mechanism). The suggested LSND effect {\it vanishes}. The neutrino mass spectrum is parametrized. In extreme cases, the spectrum may be either nearly degenerate or hierarchical (with hierarchical mass-squared differences in both cases). Always $ m_1 = -m_4 $, $m_2 = -m_5 $, $m_3 = -m_6 $ and, as can be deduced in our model from neutrino data, they are light.

\vfill\eject 

\vspace{0.2cm}

\ni {\bf 2. The model.} Let us assume in Eq. (12) that [8] 

\begin{equation} 
M^{(L)}  =  {\stackrel{0}{m}} \left( \begin{array}{ccc} 1 & 0 & 0 \\ 0 & 1 & 0 \\ 0 & 0 & 1 \end{array} \right) = - M^{(R)} 
\end{equation}

\ni and

\begin{eqnarray} 
M^{(D)} & = &  {\stackrel{0}{m}}\; U^{(3)} \left( \begin{array}{ccc} \tan 2\theta_{14} & 0 & 0 \\ 0 & \tan 2\theta_{25} & 0 \\ 0 & 0 & \tan 2\theta_{36} \end{array} \right) \nonumber \\ & = & {\stackrel{0}{m}}
 \left( \begin{array}{ccc} c_{12} \tan 2\theta_{14} & s_{12} \tan 2\theta_{25}  & 0 \\ -s_{12}c_{23} \tan 2\theta_{14} & c_{12} c_{23} \tan 2\theta_{25} & s_{23} \tan 2\theta_{36} \\ s_{12} s_{23} \tan 2\theta_{14} & -c_{12} s_{23} \tan 2\theta_{25} & c_{23} \tan 2\theta_{36} \end{array} \right) \,, 
\end{eqnarray}

\ni where ${\stackrel{0}{m}}\, > 0$ is a mass scale and $\tan 2\theta_{ij}\;(ij = 14,25,36)$ denote three dimensionless parameters, while $U^{(3)}$ stands for the previous $3\times 3$ mixing matrix given in Eq. (1). Thus, the Dirac component $M^{(D)}$ of the overall neutrino mass matrix $M$ is here equal to a diagonal, potentially hierarchical structure, deformed by the popular, nearly bimaximal mixing matrix $U^{(3)}$. Evidently, in this $6\times 6$ model, $ M^* = M$ and $M^T = M$. Hence, the possible CP violation is ignored.

As can be shown, the $6\times 6$ diagonalizing matrix $U$ for the overall $6\times 6$ mass matrix $M$ defined in Eqs. (12), (13) and (14),

\begin{equation} 
U^\dagger M U = {\rm diag}(m_1\,,\,m_2\,,\,m_3\,,\,m_4\,,\,m_5\,,\,m_6)\;,
\end{equation} 

\ni gets the form

\begin{equation}
U = {\stackrel{1}{U}}{\stackrel{0}{U}}\;,\;{\stackrel{1}{U}} =  \left( \begin{array}{cc} U^{(3)} & 0^{(3)}  \\ 0^{(3)}  & 1^{(3)} \end{array} \right) \;,\; {\stackrel{0}{U}} = \left( \begin{array}{cc} C^{(3)} & -S^{(3)} \\ S^{(3)} & C^{(3)} \end{array} \right)
\end{equation} 

\ni with $ U^{(3)}$ as given in Eq. (1) and

\begin{equation} 
1^{(3)} =  \left( \begin{array}{ccc} 1 & 0 & 0 \\ 0 & 1 & 0 \\ 0 & 0 & 1 \end{array} \right) \;,\; C^{(3)} =  \left( \begin{array}{ccc} c_{14} & 0 & 0 \\ 0 & c_{25} & 0 \\ 0 & 0 & c_{36} \end{array} \right)\;\; ,\;\;  S^{(3)} =  \left( \begin{array}{ccc} s_{14} & 0 & 0 \\ 0 & s_{25} & 0 \\ 0 & 0 & s_{36} \end{array} \right) \;,
\end{equation} 

\ni while the neutrino mass spectrum is

\begin{equation}
m_{i,j} = \pm {\stackrel{0}{m}}\sqrt{1 + \tan^2 2\theta_{i j}} 
\end{equation}

\ni ($i j = 14\,,\,25\,,\,36$), implying the equalities

\begin{equation}
\left(c^2_{i j} - s^2_{i j}\right) m_{i,j} = \pm {\stackrel{0}{m}}\;.
\end{equation}

\ni Evidently, $U^* = U$ and $U^\dagger = U^T$. 

The easiest way to prove the statement expressed in Eqs. (16), (1) and (17) is to start with the diagonalizing matrix $U = \left(U_{\alpha i} \right)$ defined in these equations, and then to show by applying the formula $M_{\alpha \beta} = \sum_i U_{\alpha i} m_i U^*_{\beta i}\,$ that the mass matrix $M = \left(M_{\alpha \beta} \right)$ is given in Eqs. (12), (13) and (14), if the mass spectrum $m_1\,,\, m_2\,,\, m_3\,,\, m_4\,,\, m_5\,,\, m_6$ is taken in the form (18) or (19). Now, $\alpha = e\,,\,\mu\,,\,\tau\,,\, e_s \,,\,\mu_s \,,\,\tau_s $ and $ i = 1,2,3,4,5,6$, where $\nu_{e_s L} \equiv \left( \nu_{e R}\right)^c$, $\nu_{\mu_s L} \equiv \left( \nu_{\mu R}\right)^c$, $\nu_{\tau_s L} \equiv \left( \nu_{\tau R}\right)^c$ (see Eq. (11) which can be rewritten as $- {\cal L}_{\rm mass} = \frac{1}{2}\sum_{\alpha \beta} \overline{(\nu_{\alpha L})^c} M_{\alpha \beta} \nu_{\beta L} + {\rm h.\,c.}$).

It may be interesting to note that the $6\times 6$ mass matrix $M$ defined in Eqs. (12), (13) and (14) can be presented as the unitary transform $ M = {\stackrel{1}{U}}{\stackrel{0}{M}} \stackrel{1}{U}\!^\dagger $ of the new simpler $6\times 6$ mass matrix
 
\begin{equation}
\stackrel{0}{M} = \left( \begin{array}{cc} \stackrel{0}{m} 1^{(3)} & \stackrel{0}{M}\!^{(D)} \\ \stackrel{0}{M}\!^{(D)\,T}  & -\stackrel{0}{m} 1^{(3)} \end{array} \right)\;{\rm with}\;\stackrel{0}{M}\!^{(D)}  = \,\stackrel{0}{m}\,{\rm diag}(\tan 2\theta_{14},\tan 2\theta_{25},\tan 2\theta_{36}) \;,
\end{equation}

\ni where $\stackrel{1}{U} =$ diag$\left( U^{(3)}, 1^{(3)} \right)$ [see Eqs. (16) and (17)]. Then, writing $\nu_{\alpha L} = \sum_\beta {\stackrel{1}{U}}_{\alpha \beta} \stackrel{0}{\nu}_{\beta L}$ with $\stackrel{0}{\nu}_{\alpha L} = \sum_i {\stackrel{0}{U}}_{\alpha i} \nu_{i L}$, the mass term (11) and charged-current term in the Lagrangian can be presented as follows:

\begin{eqnarray} 
- {\cal L}_{\rm mass} & = & \frac{1}{2} \sum_{\alpha \beta} \overline{(\nu_{\alpha L})^c} M_{\alpha \beta} {\nu_{\beta L}} +  {\rm h.\,c.} = \frac{1}{2} \sum_{\alpha \beta} (\overline{{\stackrel{0}{\nu}}_{\alpha L} )^{\!c}}\!{\stackrel{0}{M}}_{\alpha \beta} {\stackrel{0}{\nu}}_{\beta L} + {\rm h.\,c.}  \nonumber \\ & = &  \frac{1}{2} \sum_i \overline{(\nu_{i L})^c} m_i {\nu_{i L}} +  {\rm h.\,c.} 
\end{eqnarray} 

\ni $(\alpha, \beta = e, \mu, \tau, e_s, \mu_s, \tau_s \;,\; i = 1,2,3,4,5,6)$ and

\begin{eqnarray} 
- {\cal L}_{\rm CC} & = & \frac{g}{\sqrt{2}} \sum_{\alpha = e,\mu,\tau} \overline{\nu_{\alpha L}} \,  \gamma^\mu e^-_\alpha \,W^+_\mu +  {\rm h.\,c.} = \frac{g}{\sqrt{2}} \sum_{\alpha =e,\mu,\tau} \sum_{\beta =e,\mu,\tau} \overline{{\stackrel{0}{\nu} }_{\beta L}}\,\stackrel{1}{U}\!^*_{\alpha \beta} \,\gamma^\mu e^-_\alpha\, W^+_\mu + {\rm h.\,c.} \nonumber \\
 & = &\frac{g}{\sqrt{2}} \sum_{\alpha = e,\mu,\tau} \sum_{i} \overline{\nu_{i L}} \, U^*_{\alpha i} \gamma^\mu e^-_\alpha \,W^+_\mu +  {\rm h.\,c.}   
\end{eqnarray} 

\ni ($i = 1,2,3,4,5,6)$, where $e^-_\alpha = e^-\,,\,\mu^-\,,\,\tau^- \;(\alpha = e,\mu,\tau)$ and $ g = e/\sin \theta_W$. Thus, there are two different forms of CC coupling, involving $\nu_{\alpha L}$ neutrinos with nearly bimaximally mixed Dirac mass matrix $M^{(D)} = U^{(3)}{\stackrel{0}{M}}\!^{(D)}$ or ${\stackrel{0}{\nu}}_{\alpha L}$ neutrinos with hierarchical Dirac mass matrix ${\stackrel{0}{M}}\!^{(D)}$. Similarly, there are two different forms of neutral-current term ${\cal L}_{\rm NC}$ involving $\nu_{\alpha L}$ or ${\stackrel{0}{\nu}}_{\alpha L}$:

\begin{eqnarray} 
- {\cal L}_{\rm NC} & = & \frac{g}{2\cos\theta_W} \sum_{\alpha = e,\mu,\tau} \overline{\nu_{\alpha L}} \,\gamma^\mu \nu_{\alpha L} \,Z_\mu  \nonumber \\ & = & \frac{g}{2\cos\theta_W} \sum_{\alpha =e,\mu,\tau} \overline{{\stackrel{0}{\nu}}_{\alpha L}}\,\gamma^\mu {\stackrel{0}{\nu}}_{\alpha L} Z_\mu \;.  
\end{eqnarray} 

\ni In Eqs. (22) and (23)  the block-diagonal form ${\stackrel{1}{U}} =$ diag$\left( U^{(3)}, 1^{(3)} \right)$ was applied. Note that both mass matrices $M$ and ${\stackrel{0}{M}}$ develop the same eigenvalues $m_1 = - m_4\,,\, m_2 = - m_5\,,\, m_3 = - m_6$ given in Eq. (18). In fact, Eq. (15) implies that
 
\begin{equation}
{\stackrel{0}{U}}\,\!^\dagger {\stackrel{0}{M}} {\stackrel{0}{U}} = {\rm diag}(m_1\,,\, m_2\,,\, m_3\,,\, m_4\,,\, m_5\,,\, m_6)\;.
\end{equation}

\ni We can see that, in principle, one can operate with fields $\stackrel{0}{\nu}_{\alpha L}$ to describe by them the experimental flavor neutrinos, $\nu_{\alpha L} = \sum_\beta {\stackrel{1}{U}}_{\alpha \beta} {\stackrel{0}{\nu}}_{\beta L}$, using then the second forms (22) and (23) of ${\cal L}_{\rm CC}$ and ${\cal L}_{\rm NC}$. However, their transformation by means of $
{\stackrel{1}{U}}$ is necessary.

In the flavor representation, where the charged-lepton $3\times 3$ mass matrix is diagonal, the $6\times 6$ diagonalizing matrix $U$ is at the same time the neutrino $6\times 6$ mixing matrix.  Then, the unitary transformation

\begin{equation} 
\nu_{\alpha L}  = \sum_i U_{\alpha i}  \nu_{i L} \;,\; U^\dagger U = 1 = U U^\dagger
\end{equation} 

\ni holds with the $\alpha$ and $i$ indices running over their six values.

The rate for the neutrinoless double $\beta $ decay is proportional to $m^2_{e e}$, where now 

\begin{eqnarray} 
m_{e e} \equiv |\sum_i U^2_{e i} m_i| & = & | \left(U^2_{e 1} - U^2_{e 4}\right)m_1+ \left(U^2_{e 2} - U^2_{e 5}\right)m_2 + \left(U^2_{e 3} - U^2_{e 6}\right)m_3| \nonumber \\ & = & | c^2_{12}\left( c^2_{14} - s^2_{14} \right) m_1+ s^2_{12}\left( c^2_{25} - s^2_{25} \right) m_2| = \stackrel{0}{m} 
\end{eqnarray}

\ni due to Eqs. (16), (1), (17) and (19). Hence, ${\stackrel{0}{m}}\,{\stackrel{<}{\sim}}\, (0.35 - 1)$ eV is light if the suggested experimentally upper limit of $ m_{e e}$ is used. 

Making use of the neutrino oscillation formulae as given in Eq. (7) but now with $\alpha = e, \mu, \tau, e_s , \mu_s, \tau_s$ and $i = 1,2,3,4,5,6 $, we obtain the oscillation probabilities $P(\nu_\alpha \rightarrow \nu_\beta)$ for $\alpha = e, \mu, \tau$ {\it identical} with those in Eq. (9) and (10). In addition, we get $P(\nu_\alpha \rightarrow \nu_{\beta_s}) = 0$ and
$P(\nu_{\alpha_s} \rightarrow \nu_{\beta_s}) = \delta_{\beta_s \alpha_s}$ for $\alpha = e, \mu, \tau $ and $\alpha_s, \beta_s = e_s, \mu_s, \tau_s $, showing that in our $6\times 6$ model of neutrino texture there are {\it no oscillations} of (conventional) sterile neutrinos which, therefore, are strictly {\it decoupled}. In this argument, we make use of the mass-squared degeneracy relations $m^2_1 =m^2_4\,,\,m^2_2 = m^2_5\,,\,m^2_3 = m^2_6$ following from the mass spectrum (18).

From the neutrino mass spectrum (18) we infer that

\begin{eqnarray}  
\Delta m^2_{21} & = & \stackrel{0}{m}\!^2 \left(\tan^2 2\theta_{25} - \tan^2  2\theta_{14}\right) \;,\nonumber \\ \Delta m^2_{32} & = & \stackrel{0}{m}\!^2  \left(\tan^2 2\theta_{36} - \tan^2  2\theta_{25}\right) \nonumber \;, \\ \Delta m^2_{31} & = & \stackrel{0}{m}\!^2  \left(\tan^2 2\theta_{36} - \tan^2 2\theta_{14}\right) \;, 
\end{eqnarray}

\ni where experimental estimations for solar $\nu_e$'s and atmospheric $\nu_\mu$'s are

\begin{equation}  
\Delta m^2_{21} = 5\times 10^{-5}\;{\rm eV}^2 \ll \Delta m^2_{32} = 2.5\times 10^{-3}\;{\rm eV}^2 \end {equation}  

\ni if $\Delta m^2_{31} \simeq \Delta m^2_{32}$. This implies that $\tan^2 2\theta_{14} < \tan^2  2\theta_{25} \ll \tan^2 2\theta_{36}$, where $(\tan^2  2\theta_{25} - \tan^2 2\theta_{14})/\tan^2 2\theta_{36} \simeq \Delta m^2_{21}/\Delta m^2_{32} \sim 0.02$. Thus, for the neutrino mass spectrum there are, in particular, two opposite extreme options: the {\it nearly degenerate} spectrum $ m_1 \simeq m_2 \simeq m_3$, where $\tan^2 2\theta_{ij} \ll 1$ implying light $ m_i \simeq {\stackrel{0}{m}}\,{\stackrel{<}{\sim}}\, (0.35 - 1)$ eV, and the {\it hierarchical} spectrum $ m_1 < m_2 \ll m_3$, where $\tan^2 2\theta_{ij} \gg 1$ implying $ m_i \simeq {\stackrel{0}{m}} |\tan 2\theta_{ij}|$, also light since $m^2_3 \simeq \Delta m^2_{32} \sim 2.5\times 10^{-3}\,{\rm eV}^2\;( ij = 14,25,36)$. In both options, the mass-squared differences are hierarchical: $\Delta m^2_{21} \ll \Delta m^2_{32} \simeq \Delta m^2_{31}$. In the first option, $M^{(L)}$ and $M^{(R)}$ dominate over $M^{(D)}$, as can be seen from Eqs. (13) and (14). Inversely, in the second option, $M^{(L)}$ and $M^{(R)}$ are dominated by $M^{(D)}$, thus we have to do in this option with the pseudo-Dirac neutrinos [9] (of a specific sort). The first option is favored, if the actual ${\stackrel{0}{m}} = m_{e e}$ lies near its present upper limit. 

\vspace{0.3cm}

\ni {\bf 3. Conclusions.} In this note, a new explicit model of neutrino texture was presented, where in the overall $6\times 6$ mass matrix M the lefthanded and righthanded Majorana $3\times 3$ mass matrices, $M^{(L)}$ and $M^{(R)}$, are diagonal with equal entries of opposite sign, while the Dirac $3\times 3$ mass matrix $M^{(D)}$ is given as a diagonal hierarchical structure deformed by the popular nearly bimaximal $3\times 3$ mixing matrix. The neutrino $6\times 6$ mixing matrix $U$ and neutrino mass spectrum are found explicitly. Then, all neutrino masses are light and $m_1 = - m_4\,,\, m_2 = - m_5\,,\, m_3 = - m_6$, where  $m_1 \leq m_2 \leq m_3$. If, in particular, $M^{(L)}$ and $M^{(R)}$ dominate over $M^{(D)}$, then $m_1 \simeq m_2 \simeq m_3$. Inversely,  if $M^{(L)}$ and $M^{(R)}$ are dominated by $M^{(D)}$, then $m_1< m_2 \ll m_3$. But, always $\Delta m^2_{21} \ll \Delta m^2_{32} \simeq \Delta m^2_{31}$ (due to experimental estimations for solar $\nu_e$'s and atmospheric $\nu_\mu$'s). In the second option, neutrinos are pseudo-Dirac particles (of a specific sort).

The resulting neutrino oscillation formulae are {\it identical} with those working in the effective texture of three active neutrinos, based on the popular nearly bimaximal mixing matrix. Three (conventional) sterile neutrinos {\it do not oscillate} and so, are strictly {\it decoupled}. The suggested LSND effect for accelerator $\bar{\nu}_\mu$'s (and $\nu_\mu$'s) {\it vanishes}. The not observed Chooz oscillation effect for reactor $\bar{\nu}_e$'s is consistently negligible.

One may try to speculate that the Dirac component ${\stackrel{0}{M}}\,\!^{(D)}$ of ${\stackrel{0}{M}} $ as defined in Eq. (20) (before the conjectured deformation $M^{(D)} = U^{(3)}\, {\stackrel{0}{M}} \,\!^{(D)}$ by the nearly bimaximal mixing matrix $U^{(3)}$ is performed) ought to display a hierarchical structure {\it similar} to that of $3\times 3$ mass matrices for charged leptons and quarks [10] which, of course, are of Dirac type. In this sense, $ {\stackrel{0}{\nu}}_{\alpha L} = \sum_i {\stackrel{0}{U}}_{\alpha i} \nu_{i L}$ rather than $ {\nu}_{\alpha L} = \sum_i {U}_{\alpha i}{\nu}_{i L}$ are neutrino analogues of the familiar CKM transforms of down-quark mass fields. Thus, $\stackrel{0}{U}$ rather than $U$ is a lepton analogue of the CKM mixing matrix for quarks [see the first and second form of Eq. (22)]. Of course, the experimental flavor neutrino fields are ${\nu}_{\alpha L} = \sum_\beta {\stackrel{1}{U}}_{\alpha \beta} \stackrel{0}{\nu}_{\beta L}$, and only   by means of this unitary transformation the nearly bimaximal mixing (absent for ${ \stackrel{0}{\nu}}_{\alpha L}$ as well as for charged leptons and quarks) works.

\vfill\eject

~~~~
\vspace{0.5cm}

{\centerline{\bf References}}

\vspace{0.45cm}

{\everypar={\hangindent=0.6truecm}
\parindent=0pt\frenchspacing

{\everypar={\hangindent=0.6truecm}
\parindent=0pt\frenchspacing

~[1]~{\it Cf. e.g.} Z. Xing, {\it Phys. Rev.} {\bf D 61}, 057301 (2000); and references therein.

\vspace{0.2cm}

~[2]~Z. Maki, M. Nakagawa and S. Sakata, {\it Prog. Theor. Phys.} {\bf 28}, 870 (1962).

\vspace{0.2cm}

~[3]~For a recent review {\it cf.} M.C. Gonzalez-Garcia and Y.~Nir, {\tt hep--ph/0202056}; and references therein.

\vspace{0.2cm}

~[4]~{\it Cf. e.g.} M. Frigerio and A.Yu. Smirnov, {\tt hep--ph/0202247}.

\vspace{0.2cm}

~[5]~V. Barger, D. Marfatia, K. Whisnant and B.P.~Wood, {\tt hep--ph/0204253}; J.N.~Bahcall, M.C.~Gonzalez--Garcia and C. Pe\~{n}a--Garay, {\tt hep--ph/0204314v2}; and references therein.

\vspace{0.2cm}

~[6]~S. Fukuda {\it et al.}, {\it Phys. Rev. Lett.} {\bf 85}, 3999 (2000); and references therein. 

\vspace{0.2cm}

~[7]~M. Gell-Mann, P. Ramond and R.~Slansky, in  {\it Supergravity}, edited by F.~van Nieuwenhuizen and D.~Freedman, North Holland, 1979; T.~Yanagida, Proc. of the {\it Workshop on Unified Theory and the Baryon Number in the Universe}, KEK, Japan, 1979; R.N.~Mohapatra and G.~Senjanovi\'{c}, {\it Phys. Rev. Lett.} {\bf 44}, 912 (1980).

\vspace{0.2cm}

~[8]~{\it Cf.} W. Kr\'{o}likowski, {\tt hep--ph/0201186}.

\vspace{0.2cm}

~[9]~{\it Cf. e.g.} W. Kr\'{o}likowski, {\it Acta Phys. Pol.} {\bf B 30}, 663 (2000); and references therein. 

\vspace{0.2cm}

[10] W. Kr\'{o}likowski, {\tt hep--ph/0201004v2}.

\vfill\eject

\end{document}